\documentclass[12pt]{iopart}

\newcommand{\up}{\uparrow}
\newcommand{\dn}{\downarrow}
\newcommand{\non}{\nonumber}

\begin{document}

\title[Renormalization of transition matrix elements
of particle number operators]
{Renormalization of transition matrix elements
of particle number operators
due to strong electron correlation}

\author{Noboru Fukushima}

\address{Motomachi 13-23, Sanjo, Niigata, 955-0072 Japan}
\ead{noboru.fukushima@gmail.com}
\begin{abstract}
Renormalization of non-magnetic and magnetic impurities due to electron
double occupancy prohibition is derived analytically by an improved
Gutzwiller approximation.
Non-magnetic impurities are effectively weakened by the same
renormalization factor as that for the hopping amplitude, whereas
magnetic impurities are strengthened by the square root of the
spin-exchange renormalization factor, in contrast to results by the
conventional Gutzwiller approximation.
We demonstrate it by showing that transition matrix elements of number
operators between assumed excited states and between an assumed ground
state and excited states are renormalized differently than diagonal
matrix elements.
Deviation from such simple renormalization with a factor is also discussed.
In addition, as related calculation, we correct an error in treatment of
renormalization of charge interaction in the literature. Namely, terms
from the second order of the transition matrix elements are strongly suppressed.
Since all these results do not depend on the signs of impurity
potential or charge interaction parameter, they are valid both in attractive
and repulsive cases.
\end{abstract}

\maketitle
\section{Introduction}
In this paper, we discuss renormalization of impurities due to strong
electron correlation. Such renormalization may be intuitive in the case
of the Hubbard model, where each site has onsite electron
repulsion. Namely, sites with higher potential energy have lower
electron occupancy, and consequently have less chance of double
occupancy.  Then, the total energy loss from the impurity potential and
the repulsive interaction should be more uniform than in the system
without the electron correlation; we can call it renormalization of
impurities. However, when the repulsion is very strong, we need to
consider much smaller energy scales. That is, if double occupancy does
not occur, the above argument cannot be applied, and thus
renormalization of impurities within the lower Hubbard band is not so
trivial.

When electron double occupancy is prohibited at every site, a system
with quite densely packed electrons has a good chance to have one
electron with spin up or down at each site.
If the system has a tendency toward phase separation, small perturbation
by an impurity may produce a large effect to separate a system into
hole-rich regions and electron-rich regions; it may appear in close
vicinity of the half filling in the $t$-$J$--type models, where effective
hopping is negligibly small compared to effective exchange interaction.
In contrast, what we focus on in this paper are systems not that close
to the half filling or systems with relatively weak exchange
interaction. Then, electrons are more mobile.
Near the half filling, since there is little freedom left to change
charge distribution and sudden spatial change of particle number
distribution around impurities is not favorable for the kinetic energy,
non-magnetic impurity potentials may have little effect on
low-energy eigenstates and only shift their eigenenergies quite
uniformly.
In other words, impurity potentials can be renormalized by electron
correlation even within the lower Hubbard band.

In previous papers \cite{FukushimaSNS,Fukushima09}, such renormalization
of non-magnetic impurity potentials was investigated numerically.  That
is, (i) to estimate perturbation from an impurity potential, the
variational Monte Carlo method was applied to calculation of its matrix
elements with respect to assumed excited states in the uniform systems;
(ii) inhomogeneous systems with an impurity or impurities were
investigated by a Bogoliubov-de Gennes equation with the
double-occupancy prohibition treated by a kind of mean-field
approximation called the Gutzwiller approximation (GA) generalized to
inhomogeneous systems.

Both of (i) and (ii) manifested strong renormalization of the impurity
potential, and its renormalization factor (ratio between corresponding
quantities in systems with and without the double-occupancy prohibition)
seems approximately proportional to $g^t \equiv {2x}/{(1+x)}$, which is the
renormalization factor of hopping amplitude obtained by the GA as a
function of hole concentration $x$.
Since the double-occupancy prohibition inhibits hopping, $g^t$ is less
than unity and goes to zero as $x\rightarrow 0$.
To explain the impurity renormalization factor close to $g^t$, we
pointed out the similarity between the impurity potential and the
hopping in the real space, i.e., the Fourier-transformed impurity
potential has the form of hopping in the $k$-space. If electrons are
densely packed in the lattice, it must be difficult even in the
$k$-space to hop from $k$ to a different $k'$.

However, it is a speculation and may not be trivial because the double
occupancy is prohibited in the real space rather than in the $k$-space.
In addition, we do not really know how general the numerical results
are because the calculation was done only for limited parameter sets.
To complement this argument, an analytic approximation is adopted in this paper,
namely, (i) is redone using the GA to derive dominant $g^t$ dependence
and deviation from $g^t$ explicitly. 
In fact, however, the conventional GA \cite{Gutzwiller65, FCZhang88}
fails to derive this renormalization.  It compares mean weights of
configurations relevant to operators of interest with and without the
electron repulsion in calculating the renormalization factors. Then, the
renormalization factor for the particle number operators is actually
unity, i.e., they are not renormalized.
The spin rotation invariant slave-boson mean-field theory \cite{TLi89} is
known to be equivalent to the conventional GA; the saddle-point
approximated boson fields play a role of the weights in the
GA. Therefore, we speculate that it may have the same problem as the
conventional GA.  In addition, we believe that the slave-boson
mean-field theory with only one boson often used for the $t$-$J$ model
can be even less accurate because it does not yield renormalization of the
exchange interaction, which may be an artifact from the lost boson
hard-core property.

Let us recall that the GA corresponds to taking the
leading order of the Wick expansion with respect to the intersite
contractions of creation/annihilation operators
\cite{Gebhard90,Fukushima08}.  In fact, the weights of configurations in
the conventional GA are likely to be calculated with the focus only on
the lowest order; apparently it breaks
down when the lowest order vanishes or when the next lowest order is
of interest. An example is a particle number operator as shown in this paper.
Although the lowest order is the average particle number, when we discuss
transition matrix elements with excited states, this lowest order
does not contribute, and the next lowest order is relevant.
We will demonstrate that off-diagonal matrix elements between
an assumed ground state and excited states as well as between different
excited states are renormalized differently than diagonal matrix
elements.

Furthermore, by slightly modifying the non-magnetic impurity, i.e., by
subtraction between up- and down-spin particle number operators, we also
consider a simple magnetic impurity.  In this case, the direction of the
renormalization is reversed, namely, the impurity is strengthened by the
electron correlation in contrast to the non-magnetic impurity.  It must
be physically reasonable because electron repulsion increases single
occupancy.
As calculation related to the non-magnetic impurity, renormalization of
charge interaction is discussed to correct an error in
its treatment in the literature. That is, terms relevant to the
mean-field approximation are actually the second order of the transition
matrix elements, and they are weakened by very small renormalization
factor $(g^t)^2$ although treated usually as not being renormalized.

\section{Model}

What we have in mind is $t$-$J$--type models with impurities, namely,
\begin{equation}
H \equiv 
P_{\rm G}\left[- \sum_{i,j,\sigma} t_{ij}  c_{i\sigma}^\dagger
           c_{j\sigma}
+ \sum_{i,j} J_{ij}
\left( {\bf S}_i\cdot {\bf S}_j -\frac14 \hat{n}_i \hat{n}_j \right)
+ H_{\rm imp}
 \right] P_{\rm G}
,
\label{eq:Hamiltonian}
\end{equation}
where $c_{i\sigma}^\dagger$ ($c_{i\sigma}$) is the creation
(annihilation) operator of the electron with site $i$ and spin $\sigma$,
and ${\bf S}_i$ is the spin operator at site $i$. In addition,
\begin{equation}
\hat{n}_i\equiv\hat{n}_{i\up}+\hat{n}_{i\dn},
\qquad
\hat{n}_{i\sigma} \equiv c_{i\sigma}^\dagger c_{i\sigma}
.
\end{equation}
Gutzwiller projection operator
$
 P_{\rm G} \equiv \prod_i (1-\hat{n}_{i\uparrow} \hat{n}_{i\downarrow})
$
prohibits electron double occupancy at each site 
and represents strong Coulomb repulsion.
In this paper, we do not use any explicit form of $t_{ij}$ and $J_{ij}$
although they are implicitly included in assumed variational
ground/excited states.  Our main focus here is on the impurity term
$H_{\rm imp}$.

In sections~\ref{sec:nonmag}, \ref{sec:nonmagcorr} and \ref{sec:nonmaggen},
our target is renormalization of a single non-magnetic
$\delta$-function impurity potential located at $i=I$,
\begin{equation}
 H_{\rm imp} = V_I \hat{n}_{I}= V_I ( \hat{n}_{I\up} + \hat{n}_{I\dn} )
\label{eq:nonmagimpHamil}
.
\end{equation}
Then, in section~\ref{sec:mag}, we discuss renormalization of a simple
magnetic impurity,
\begin{equation}
 H_{\rm imp} = - h_I S^z_I = - \frac{h_I}{2} ( \hat{n}_{I\up} - \hat{n}_{I\dn} )
.
\label{eq:magimpHam}
\end{equation}
In addition, the focus in section~\ref{sec:chargeint} is not on $H_{\rm
imp}$ but on charge interaction $\hat{n}_i \hat{n}_j$ in
Hamiltonian (\ref{eq:Hamiltonian}).

\section{Non-magnetic impurity renormalization}
\label{sec:nonmag}

Let us start from a uniform system without impurities.
A basic idea of variational theories is that the ground state of the
$t$-$J$--type models may be something similar to the BCS
superconducting state
\begin{equation}
|\Psi_0 \rangle \equiv
\prod_k \left( u_k + v_k c_{k\up}^\dagger c_{-k\dn}^\dagger \right) |0\rangle
,
\end{equation}
but somewhat modified by the electron correlation. Simple variational
wave functions adopted by most of analytic theories have a form of
$P_{\rm G}|\Psi_0 \rangle$ with something to control the particle
number.
One way to control it is to use projection $P_N$ to fixed
particle number $N$.\footnote{
Many different $|\Psi_0\rangle$ correspond to $|\Psi\rangle$ under the
projections. For example, $\exp(\lambda \hat{N})$ with $\hat{N}$ the
total particle number operator is constant under $P_N$,
and thus $\exp(\lambda \hat{N})|\Psi_0\rangle$ is equivalent to
$|\Psi_0\rangle$.
}
Another is to attach fugacity factors to the projector, namely,
\begin{equation}
|\Psi\rangle \equiv P |\Psi_0 \rangle
, \qquad
P \equiv  \prod_i P_i
, \qquad
P_i \equiv
 \lambda_{i\uparrow}^{\frac12\hat{n}_{i\uparrow} }
 \lambda_{i\downarrow}^{\frac12 \hat{n}_{i\downarrow}}
 (1-\hat{n}_{i\uparrow} \hat{n}_{i\downarrow})
.
\end{equation}
The latter is adopted in this paper.  The reason to control the particle
number is that $P_{\rm G}$ changes the average particle number of $|\Psi_0
\rangle$ because states with a larger particle number have more chance
to be projected out \cite{Edegger05}. Since the GA relates 
expectation values before and after the projection,
\begin{equation}
\langle \hat{O} \rangle_0 \equiv
\langle \Psi_0| \hat{O}|\Psi_0\rangle
,
\qquad
\langle \hat{O} \rangle \equiv
\frac {\langle \Psi|
 \hat{O}|\Psi\rangle}{\langle \Psi |\Psi\rangle}
,
\end{equation}
for some operator $\hat{O}$, usually it is not convenient if $|\Psi\rangle$ and
$|\Psi_0\rangle$ are totally different, e.g., if $|\Psi_0\rangle$ has a
more-than-half filled electron band.\footnote{ The variational Monte
Carlo method does not have such restriction.  For example, local
magnetic moments before and after the projection are different in
general, and the chemical potential in a variational mean-field
Hamiltonian is a variational parameter under $P_N$ rather than a
parameter to control the particle number. }
Although our main interest here is perturbation from the uniform state,
most of derivation in this paper is valid also for inhomogeneous
systems, and thus we prefer to keep general expressions with site and
spin indices throughout the paper, e.g.,
\begin{equation}
n_{i\sigma} \equiv \langle \hat{n}_{i\sigma} \rangle_0,\qquad
n_i \equiv \langle \hat{n}_i \rangle_0 = n_{i\uparrow} + n_{i\downarrow}
.
\end{equation}
However, we use
$ 0 = \langle c_{i\up}^{\dagger}c_{i\dn}^{\dagger}\rangle_0
=\langle c_{i\sigma}^\dagger c_{j\sigma}^\dagger \rangle_0=
\langle c_{i\up}^\dagger c_{j\dn} \rangle_0$
to avoid making formulas too lengthy.

Although choice of the fugacity factors is not unique especially in
inhomogeneous systems \cite{Fukushima08}, yet it is convenient to
define
\begin{equation}
\lambda_{i\sigma} \equiv \frac{1-n_{i\sigma}}
{ 1-n_{i} }
,
\label{eq:deflambdai}
\end{equation}
because it satisfies
\begin{equation}
\langle \hat{n}_{i\sigma} \rangle \approx \langle \hat{n}_{i\sigma} \rangle_0
,
\label{eq:localconst}
\end{equation}
for any $i$ and $\sigma$  \cite{Gebhard90,Fukushima08},
neglecting terms of the ``fourth order''.
Here, and throughout this paper, if not specified,
``$n$-th order'' represents
$n$-th order with respect to intersite contractions
such as $\langle c_{i\sigma}^\dagger c_{j\sigma} \rangle_0$ and
$\langle c_{i\dn} c_{j\up} \rangle_0$ with $i\neq j$.
Note that $\langle \hat{O} \rangle $ of any $\hat{O}$
can be in principle calculated by the Wick theorem, which yields
many such intersite contractions.
High order terms may be neglected by recalling that onsite
contractions are larger than intersite contractions.
The GA corresponds to taking the leading order only, e.g.,
\begin{equation}
\langle P^2 \rangle_0
\approx
\prod_i \langle P_i^2 \rangle_0
,
\label{eq:denomiGutz}
\end{equation}
\begin{eqnarray}
\langle P_i^2 \rangle_0 &=& (1-n_{i\up}) (1-n_{i\dn})
+ \lambda_{i\up} n_{i\up}(1-n_{i\dn}) + \lambda_{i\dn} n_{i\dn}(1-n_{i\up}) 
\nonumber \\
&=&
 \frac{ (1-n_{i\up}) (1-n_{i\dn}) }{ 1-n_{i} }
.
\end{eqnarray}
The terms neglected in the approximation in
(\ref{eq:denomiGutz}) are of the fourth order because the second
order terms cancel out when $\lambda_{i\sigma}$ is defined as
(\ref{eq:deflambdai}) \cite{Gebhard90,Fukushima08}.
Let us show it explicitly with a notation to treat $c^\dagger$ and $c$
together,
\begin{equation}
 c_{i\sigma}^{+} \equiv c_{i\sigma}^{\dagger}, \qquad
 c_{i\sigma}^{-} \equiv c_{i\sigma}
,
\end{equation}
by considering contractions between $P_i^2$ and operators at
some site(s) $j,j'\ne i$,
\begin{eqnarray}
\fl
\langle P_i^2  c_{j'\sigma'}^{\tau'} c_{j\sigma}^{\tau} \rangle_0 =
\langle P_i^2 \rangle_0
\langle c_{j'\sigma'}^{\tau'} c_{j\sigma}^{\tau} \rangle_0
\non \\
\fl
\qquad
 +
\left(-
\langle c_{i\up} c_{j'\sigma'}^{\tau'}  \rangle_0
\langle c_{i\up}^{\dagger} c_{j\sigma}^{\tau} \rangle_0
+
\langle c_{i\up} c_{j\sigma}^{\tau}  \rangle_0
\langle c_{i\up}^{\dagger} c_{j'\sigma'}^{\tau'} \rangle_0
\right)
[(1-n_{i\dn})-\lambda_{i\up}(1-n_{i\dn})+\lambda_{i\dn}n_{i\dn}]
\non \\
\fl
\qquad
+
\left(-
\langle c_{i\dn} c_{j'\sigma'}^{\tau'}  \rangle_0
\langle c_{i\dn}^{\dagger} c_{j\sigma}^{\tau} \rangle_0
+
\langle c_{i\dn} c_{j\sigma}^{\tau}  \rangle_0
\langle c_{i\dn}^{\dagger} c_{j'\sigma'}^{\tau'} \rangle_0
\right)
[(1-n_{i\up})+\lambda_{i\up}n_{i\up}-\lambda_{i\dn}(1-n_{i\up})]
\end{eqnarray}
for arbitrary $\tau$, $\tau'$, $\sigma$ and $\sigma'$. Then, the
quantities in the square brackets vanish.

We assume that $|\Psi\rangle$ is a good variational ground state, and
that the excited states are well represented by
projected quasiparticles
\begin{equation}
| k s \rangle
\equiv
\frac{
 P\gamma_{k s}^\dagger |\Psi_0 \rangle
}
{
\sqrt{
\langle \Psi_0|\gamma_{k s}
 P P
\gamma_{k s}^\dagger |\Psi_0 \rangle
}
}
\approx
\frac{
 P\gamma_{k s}^\dagger |\Psi_0 \rangle
}
{
\sqrt{
\langle P^2 \rangle_0
}
}
,
\label{eq:defks}
\end{equation}
where $\gamma_{ks}$ are quasiparticles for $|\Psi_0 \rangle$, namely,
\begin{equation}
 \gamma_{k\up}^\dagger = u_k^* c_{k\up}^\dagger  - v_k^* c_{-k\dn}, \qquad
 \gamma_{-k\dn}         = v_k c_{k\up}^\dagger + u_k c_{-k\dn}
.
\end{equation}
For the denominator of $|k s \rangle$,
we have used approximation $\langle \Psi_0|\gamma_{k\sigma}
 P^2
\gamma_{k\sigma}^\dagger |\Psi_0 \rangle
\approx
\langle P^2 \rangle_0
$ \cite{Fukushima08,Laughlin02},
and errors from this approximation are of the second order.

By switching on the impurity potential, these excited states should
be mixed by matrix elements
\begin{equation}
\frac{ V_{k',k} }{N_{\rm L}}\equiv\langle k's| \hat{n}_I |ks\rangle
\approx
\frac{
\left\langle \gamma_{k's}
P \hat{n}_{I} P
\gamma_{ks}^\dagger \right\rangle_0
}
{\langle P^2 \rangle_0}
,
\end{equation}
with $N_{\rm L}$ the number of sites.
The limit of the half filling can be exactly evaluated;
$\lambda\rightarrow\infty$, $P \hat{n}_{I} P\rightarrow PP$, and thus
$V_{k',k}/N_{\rm L}\rightarrow \langle k's|ks\rangle = \delta_{k'k}$.
According to the BCS theory,
$V_{k',k}^{\rm BCS} \equiv \langle\gamma_{k's} \hat{n}_I \gamma_{ks}^\dagger\rangle_0
= u_{k'}u_{k}^* - v_{k'} v_{k}^*$.
In the previous paper \cite{Fukushima09}, the author noted that
$V_{k',k}$ is not renormalized with the conventional GA \cite{FCZhang88}
because it originally comes from a particle number operator. However,
more careful analysis here will show that, although the {\it diagonal}
matrix elements of the particle number operators are not renormalized
[eg., see (\ref{eq:localconst})], their {\it off-diagonal} matrix
elements with respect to the projected quasiparticle excited states
are renormalized.

The Wick expansion of $\langle \gamma_{k's} P \hat{n}_{I\sigma} P
\gamma_{ks}^\dagger \rangle_0$
yields many terms, and some terms contain onsite contraction of
$\hat{n}_{I\sigma}$ at the center as $\hat{n}_{I\sigma}\rightarrow
{n}_{I\sigma}$, and the others do not. Let us separate these two
groups of terms,
\begin{equation}
 \langle  \gamma_{k's}
P \hat{n}_{I\sigma} P
 \gamma_{ks}^\dagger \rangle_0
= {n}_{I\sigma}\langle \gamma_{k's}
P^2
 \gamma_{ks}^\dagger  \rangle_0
+ \langle \gamma_{k's}
P^2 (\hat{n}_{I\sigma}-{n}_{I\sigma})
 \gamma_{ks}^\dagger  \rangle_0
.
\label{eq:twogroup}
\end{equation}
The first term is proportional to $\langle k'\up|k\up\rangle$,
and vanishes when $k\neq k'$.
Namely, we can only consider the second term.

Let us first take only $\hat{n}_{I\up}$ in the impurity potential term.
Since the GA is carried out in the real space, the $k$ representation should
be inverse Fourier transformed into the real space representation.
Namely, what we should calculate is $\langle c_{i'\sigma'}^{\tau'}
P \hat{n}_{I\up} P c_{i\sigma}^{\tau}  \rangle_0$.
Let us first take the case of $i\neq I$, $i'\neq I$ and $i\neq i'$,
which makes dominant contribution to $V_{k'k}$.
After using $P_I \hat{n}_{I\up} P_I=\lambda_{I\up}
\hat{n}_{I\up}(1-\hat{n}_{I\dn})$,
we take onsite contractions for all the sites except $i$, $i'$ and $I$
of the numerator neglecting fourth-order terms,
\begin{equation}
\frac{
 \langle c_{i'\sigma'}^{\tau'}
P \hat{n}_{I\up} P
c_{i\sigma}^{\tau}  \rangle_0
}
{
\langle P^2 \rangle_0
}
\approx
\frac{
\lambda_{I\up} \langle c_{i'\sigma'}^{\tau'} P_{i'}^2 \
\hat{n}_{I\up}(1-\hat{n}_{I\dn}) \
 P_{i}^2 c_{i\sigma}^{\tau} \rangle_0
}
{
\langle P_{i'}^2 \rangle_0 \langle P_I^2 \rangle_0\langle P_i^2 \rangle_0
}
.
\label{eq:ciPnPci1}
\end{equation}
Then, let us work on sites $i$ and $i'$,
\begin{equation}
P_i^2 c_{i\sigma}^\dagger =
\lambda_{i \sigma} (1-\hat{n}_{i\bar{\sigma}} )\, c_{i\sigma}^\dagger 
, \qquad
P_i^2 c_{i\sigma} =
\big[(1-\hat{n}_{i\bar{\sigma}} ) +
\lambda_{i\bar\sigma}  \hat{n}_{i\bar{\sigma}} \big]c_{i\sigma}
.
\label{eq:P2c}
\end{equation}
For the moment, we take the onsite contractions for $i\bar\sigma$ and
$i'\bar{\sigma'}$ neglecting intersite contractions between $I\up$ or
$I\dn$ and them; the terms neglected here are of the third order and
will be calculated in the next section.
Accordingly, using
\begin{equation}
\frac{ \lambda_{i \sigma} (1-n_{i\bar{\sigma}} )}{\langle P_i^2 \rangle_0}
=
\frac{(1-n_{i\bar{\sigma}} ) +
\lambda_{i\bar\sigma}  n_{i\bar{\sigma}} }{\langle P_i^2 \rangle_0}=1
,
\end{equation}
(\ref{eq:ciPnPci1}) can be approximated as
\begin{equation}
\frac{
 \langle c_{i'\sigma'}^{\tau'}
P \hat{n}_{I\up} P
c_{i\sigma}^{\tau}  \rangle_0
}
{
\langle P^2 \rangle_0
}
\approx
\frac{
\lambda_{I\up} \langle c_{i'\sigma'}^{\tau'}
\hat{n}_{I\up}(1-\hat{n}_{I\dn})
c_{i\sigma}^{\tau}  \rangle_0
}
{
\langle P_I^2 \rangle_0
}
.
\label{eq:nuprenoyet}
\end{equation}
It is convenient to define mean-value--subtracted operators here,
\begin{equation}
 \tilde{n}_{i\sigma} \equiv \hat{n}_{i\sigma} - {n}_{i\sigma}
.
\end{equation}
Consequently, we obtain
\begin{equation}
\frac{
 \langle c_{i'\sigma'}^{\tau'}
P \tilde{n}_{I\up}  P
c^{\tau}_{i\sigma}  \rangle_0
}
{
\langle P^2 \rangle_0
}
\approx
\langle c_{i'\sigma'}^{\tau'}\tilde{n}_{I\up} c_{i\sigma}^{\tau} \rangle_0
-\frac{ n_{I\up} }{ 1 - n_{I\dn} }
\langle c_{i'\sigma'}^{\tau'} \tilde{n}_{I\dn} c_{i\sigma}^{\tau}
\rangle_0
.
\label{eq:nuprenofin}
\end{equation}
Here, the first term and the second term in the r.h.s.\ are from the
onsite contraction of $1-\hat{n}_{I\dn}$ and $\hat{n}_{I\dn}$,
respectively; from the residual operators ($\hat{n}_{I\up}$ and
$1-\hat{n}_{I\dn}$, respectively), their mean values are subtracted to
cancel their onsite contraction.

For the moment, we neglect deviation from (\ref{eq:nuprenofin}) for
any $i$ and $i'$, which will be discussed in the next section.  Then,
it is straightforward to Fourier transform back,
\begin{equation}
\frac{
 \langle \gamma_{k's}
P \tilde{n}_{I\up} P
\gamma_{ks}^\dagger \rangle_0
}
{
\langle P^2 \rangle_0
}
\approx
\langle \gamma_{k's}
 \tilde{n}_{I\up} 
\gamma_{ks}^\dagger \rangle_0
-\frac{ n_{I\up} }{ 1 - n_{I\dn} }
\langle \gamma_{k's}
 \tilde{n}_{I\dn}
\gamma_{ks}^\dagger
\rangle_0
.
\label{eq:nuprenogamma}
\end{equation}
The formula for $\hat{n}_{I\dn}$ is obtained by exchanging $\up$ and $\dn$
at site $I$, and these formula represent that $\tilde{n}_{I\sigma}$ is
renormalized into $\tilde{n}_{I\sigma} - \tilde{n}_{I\bar\sigma}
n_{I\sigma}/(1 - n_{I\bar\sigma}) $.

In fact, the derivation above is valid also for inhomogeneous systems by
replacing $\gamma_{ks}$ with Bogoliubov quasiparticles $\gamma_\ell$.  A
difference is that the orthogonality of the Gutzwiller-projected
Bogoliubov quasiparticle states is only approximately satisfied
\cite{Fukushima08}, i.e., errors from the GA can be larger than those in
uniform systems.  The renormalization of $\hat{n}_I$ in inhomogeneous
systems is obtained by summing up $\hat{n}_{I\up}$ and $\hat{n}_{I\dn}$
for $\ell \neq \ell'$,
\begin{equation}
\frac{\langle \gamma_{\ell'} P \hat{n}_I P \gamma_{\ell}^\dagger \rangle_0 }
{\sqrt{\langle \gamma_{\ell'} P P \gamma_{\ell'}^\dagger \rangle_0
       \langle \gamma_{\ell} P P \gamma_{\ell}^\dagger \rangle_0} }
\approx
\left \langle \gamma_{\ell'}
\left(
g^t_{I\up} \tilde{n}_{I\up}
+
g^t_{I\dn}\tilde{n}_{I\dn}
\right)
\gamma_{\ell}^\dagger \right\rangle_0
,
\label{eq:gtnI}
\end{equation}
where
\begin{equation}
g^t_{i\sigma} \equiv \frac{1-n_i}{1-n_{i\sigma}}\end{equation} is the
Gutzwiller renormalization factor for the hopping amplitude.

Returning to our main target, i.e., the non-magnetic uniform system,
we can set $g^t_{Is}=g^t_{I\bar{s}}$, then
\begin{equation}
 V_{k',k} = \left\langle k's |\hat{n}_{I}|ks \right\rangle
\approx g^t_{Is}( u_{k'} u_{k}^* -  v_{k'} v_{k}^*)
=  g^t_{Is} V_{k',k}^{\rm BCS}
,
\label{eq:vkk}
\end{equation}
which is exactly the same as the speculation in the previous paper
\cite{Fukushima09} consistent with the numerical results for several $k$-points
by the variational Monte Carlo method, i.e., the renormalization factor is
close to $g^t$ and insensitive to model parameters.
The important point here may be $g^t$ appears only after summation of up
and down spins, $\hat{n}_{I\up} + \hat{n}_{I\dn}$, which is a difference
from the hopping amplitude renormalization in the real space.

According to the conventional GA \cite{FCZhang88}, what is renormalized
is an operator rather than its matrix elements, and thus diagonal and
off-diagonal matrix elements have the same renormalization factor. In
fact, however, what is renormalized should be matrix elements rather
than operators, and diagonal and off-diagonal matrix elements with
respect to excited states can have different renormalization factors as
demonstrated above.

By exactly the same procedure as above, transition matrix elements
between the variational ground state and projected two-quasiparticle
excited states can be also calculated.
Corresponding to (\ref{eq:gtnI}),
\begin{equation}
\frac{\langle \Psi_0| \gamma_{\ell}\gamma_{\ell'} P \hat{n}_I |\Psi\rangle }
{\sqrt{\langle \Psi_0| \gamma_{\ell} \gamma_{\ell'} P P
\gamma_{\ell'}^\dagger \gamma_{\ell}^\dagger |\Psi_0\rangle
       \langle \Psi | \Psi \rangle} }
\approx
\langle \Psi_0 | \gamma_{\ell} \gamma_{\ell'}
\left(
g^t_{I\up} \tilde{n}_{I\up}
+
g^t_{I\dn}\tilde{n}_{I\dn}
\right)
 |\Psi_0 \rangle
.
\label{eq:gtnIgs}
\end{equation}

\section{Corrections to the simple $g^t$ renormalization }
\label{sec:nonmagcorr}

In the cases of $i=I\neq i'$, $i'=I\neq i$ and $i=i'=I$,
we obtain formulas equivalent to (\ref{eq:nuprenofin}).
However, for $i=i'\neq I$, we have
\begin{equation}
\frac{
 \langle c_{i\sigma'}^{\tau'}
P \hat{n}_{I\up} P
c_{i\sigma}^{\tau}  \rangle_0
}
{
\langle P^2 \rangle_0
}
\approx
\frac{
\lambda_{I\up} \langle
\hat{n}_{I\up}(1-\hat{n}_{I\dn}) \
c_{i\sigma'}^{\tau'} P_{i}^2 c_{i\sigma}^{\tau} \rangle_0
}
{
 \langle P_I^2 \rangle_0\langle P_i^2 \rangle_0
}
,
\end{equation}
where $c_{i\sigma'}^{\tau'} P_{i}^2 c_{i\sigma}^{\tau} $ can be
explicitly written as
\begin{equation}
\fl
c_{i\sigma} P_i^2 c_{i\sigma}^\dagger =
\lambda_{i \sigma} (1-\hat{n}_{i\bar{\sigma}} ) (1-\hat{n}_{i\sigma})
, \qquad
c_{i\sigma}^\dagger P_i^2 c_{i\sigma} =
\big[(1-\hat{n}_{i\bar{\sigma}} ) +
\lambda_{i\bar\sigma}  \hat{n}_{i\bar{\sigma}} \big] \hat{n}_{i\sigma}
,
\end{equation}
because the other combinations of $c_{i\sigma'}^{\tau'}$ and
$c_{i\sigma}^{\tau} $ yield zero or very small quantities.
Then, although the onsite contraction of $i\bar\sigma$ with intersite
contractions between $i\sigma$ and $I$ yields a formula
equivalent to (\ref{eq:nuprenofin}),
the onsite contraction of $i\sigma$ with intersite contractions between
$i\bar\sigma$ and $I$ additionally yields the same order of
contribution. To compactly write them, let us define
\begin{equation}
\kappa_{i\sigma}^+ \equiv
- \frac{ 1 }{ 1 - n_{i\bar\sigma} }
, \qquad
\kappa_{i\sigma}^- \equiv
 \frac{ n_{i\sigma} }{ (1 - n_{i\up})(1 - n_{i\dn}) }
,
\end{equation}
as well as
\begin{equation}
\eta_{i'\sigma', i\sigma} \equiv
\langle \hat{n}_{i'\sigma'} \hat{n}_{i\sigma} \rangle_0
- {n}_{i'\sigma'} {n}_{i\sigma},
\end{equation}
which extracts only intersite contractions in
$\langle \hat{n}_{i'\sigma'} \hat{n}_{i\sigma} \rangle_0$.
Then, $\langle \hat{n}_{i'\sigma'} (1-\hat{n}_{i\sigma}) \rangle_0
- {n}_{i'\sigma'} (1-{n}_{i\sigma})= - \eta_{i'\sigma', i\sigma}$,
and $\langle (1-\hat{n}_{i'\sigma'}) (1-\hat{n}_{i\sigma}) \rangle_0
- (1-{n}_{i'\sigma'}) (1-{n}_{i\sigma})=  \eta_{i'\sigma', i\sigma}$.
More explicitly,
\begin{equation}
 \eta_{i'\sigma, i\sigma} =
- \left|\langle c_{i'\sigma}^\dagger c_{i\sigma}\rangle_0\right|^2
, \qquad
 \eta_{i'\bar\sigma, i\sigma} =
\left|
\langle c_{i'\bar\sigma}^\dagger c_{i\sigma}^\dagger \rangle_0
\right|^2
.
\end{equation}
Using these notations,
\begin{equation}
\fl
\frac{
 \langle c_{i\sigma}^{\bar\tau}
P \tilde{n}_{I\up} P
c_{i\sigma}^{\tau}  \rangle_0
}
{
\langle P^2 \rangle_0
}
\approx
\bar\tau \left(
\eta_{I\up, i \sigma}
- \frac{ n_{I\up} }{ 1 - n_{I\dn} } \eta_{I\dn, i \sigma}
\right)
+ \kappa_{i\sigma}^\tau
 \langle c_{i\sigma}^{\bar\tau} c_{i\sigma}^{\tau}  \rangle_0
\left(\eta_{I\up, i \bar\sigma}
- \frac{ n_{I\up} }{ 1 - n_{I\dn} }
 \eta_{I\dn, i \bar\sigma}
\right)
.
\label{eq:ieqipnup}
\end{equation}
By summing up $\hat{n}_{I\up}$ and $\hat{n}_{I\dn}$,
\begin{equation}
\fl
\frac{
 \langle c_{i\sigma}^{\bar\tau}
P ( \tilde{n}_{I\up}+\tilde{n}_{I\dn} ) P
c_{i\sigma}^{\tau}  \rangle_0
}
{
\langle P^2 \rangle_0
}
\approx
\bar\tau(
  g^t_{I\up} \eta_{I\up, i \sigma}
+ g^t_{I\dn} \eta_{I\dn, i \sigma}
)
+  \kappa_{i\sigma}^\tau
 \langle c_{i\sigma}^{\bar\tau} c_{i\sigma}^{\tau}  \rangle_0
\left(g^t_{I\up} \eta_{I\up, i \bar\sigma}
+ g^t_{I\dn}  \eta_{I\dn, i \bar\sigma}
\right)
.
\label{eq:ieqipnupdn}
\end{equation}

Since $i \neq i'$ occurs more often than $i=i'$, the third-order terms
neglected in the previous section for the case of $i\neq I$, $i'\neq I$,
and $i\neq i'$ may have larger contribution than
the newly derived terms above.
Such terms are derived by taking into account intersite contraction including
$i\bar\sigma$ and $i'\bar{\sigma'}$.
However, if intersite contractions are taken between $i\bar\sigma$
and $i'\bar{\sigma'}$ and the onsite contractions are taken for $I$, then
such terms do not contribute as explained around
(\ref{eq:twogroup}). 
Using the notation above, (\ref{eq:P2c}) is rewritten as
\begin{equation}
\frac{P_i^2 c_{i\sigma}^\tau}{\langle P_i^2\rangle_0} =
(1 + \kappa_{i\sigma}^\tau \tilde{n}_{i\bar{\sigma}} ) c_{i\sigma}^\tau
.
\end{equation}
Then, for $\tau'= - \tau\sigma\sigma'$
($\up,\dn$ and $+1,-1$ are used interchangeably)
,
\begin{eqnarray}
\fl 
\frac{
 \langle c_{i'\sigma'}^{\tau'}
P (\tilde{n}_{I\up}+\tilde{n}_{I\dn}) P
c_{i\sigma}^\tau  \rangle_0
}
{
\langle P^2 \rangle_0
}
 \approx
g^t_{I\up}
\langle c_{i'\sigma'}^{\tau'}
\tilde{n}_{I\up}
c_{i\sigma}^\tau  \rangle_0
+g^t_{I\dn}
\langle c_{i'\sigma'}^{\tau'}
\tilde{n}_{I\dn}
c_{i\sigma}^\tau  \rangle_0
\non \\
\fl 
\qquad
+ \langle c_{i'\sigma'}^{\tau'} c_{i\sigma}^\tau  \rangle_0
\left[
\kappa_{i\sigma}^\tau
\left(g^t_{I\up}\eta_{I\up, i \bar\sigma}
   +  g^t_{I\dn}\eta_{I\dn, i \bar\sigma}
\right)
+
\kappa_{i'\sigma'}^{\bar{\tau'}}
\left(g^t_{I\up}\eta_{I\up, i' \bar{\sigma'}}
   + g^t_{I\dn}\eta_{I\dn, i' \bar{\sigma'}}
\right)
\right]
\non \\
\fl 
\qquad
+ \sigma \kappa_{i\sigma}^\tau
\langle c_{i'\sigma'}^{\tau'} c_{i\bar\sigma}^{\bar\tau}  \rangle_0
\left( g^t_{I\up}
 \langle c_{I\up}^\dagger c_{i\bar\tau}^{\tau} \rangle_0
 \langle c_{I\up} c_{i\tau}^{\tau} \rangle_0 
   -  g^t_{I\dn}
 \langle c_{I\dn}^\dagger c_{i\tau}^{\tau} \rangle_0
 \langle c_{I\dn} c_{i\bar\tau}^{\tau} \rangle_0 
\right)
\non \\
\fl 
\qquad
+ \sigma' \kappa_{i'\sigma'}^{\bar{\tau'}}
\langle c_{i'\bar{\sigma'}}^{\bar{\tau'}}  c_{i\sigma}^{\tau}  \rangle_0
\left( g^t_{I\up}
 \langle c_{I\up}^\dagger c_{i'\bar{\tau'}}^{\tau'} \rangle_0
 \langle c_{I\up} c_{i'\tau'}^{\tau'} \rangle_0 
   -  g^t_{I\dn}
 \langle c_{I\dn}^\dagger c_{i'\tau'}^{\tau'} \rangle_0
 \langle c_{I\dn} c_{i'\bar{\tau'}}^{\tau'} \rangle_0 
\right)
.
\label{eq:highernupdn}
\end{eqnarray}

Although all the new terms in (\ref{eq:ieqipnupdn}) and
(\ref{eq:highernupdn}) contain the $g^t$ factors, they are
not so simple as (\ref{eq:nuprenofin}) and inhibit the straightforward
analytical transform back to $k$-representation.  In other words, they
cause $k$-dependence of the renormalization.
Since the ratio between the leading order and the corrections
calculated in (\ref{eq:highernupdn}) is only of the first order, the
influence from the corrections may be larger than those in the GA for the
real-space hopping amplitude, where the ratio is of the second order.

Other corrections are the terms neglected in (\ref{eq:defks}). We
expect that they only slightly change the magnitude of the leading
order, and that their contribution is probably not very important.

\section{General estimation of higher-order terms}
\label{sec:nonmaggen}

Let us estimate the other higher-order terms neglected above.  The
terms appearing in the Wick expansion can be classified into three groups
by how to take contractions of $\hat{n}_{I\sigma}
(1-\hat{n}_{I\bar\sigma})$: (i) Onsite contractions are taken both for
$I\up$ and $I\dn$. These terms do not contribute to $V_{k'k}$ as
explained around (\ref{eq:twogroup}).
(ii) If intersite contractions are taken for $I\sigma$ and the onsite
contraction is taken for $I\bar\sigma$, then $\lambda_{I\sigma}
\hat{n}_{I\sigma} (1-\hat{n}_{I\bar\sigma})/\langle P_I^2 \rangle_0$ is
reduced to $\tilde{n}_{I\sigma}$.  Doing the same for
$\lambda_{I\bar\sigma}\hat{n}_{I\bar\sigma}
(1-\hat{n}_{I\sigma})/\langle P_I^2 \rangle_0$ yields
$-\tilde{n}_{I\sigma} n_{I\bar\sigma}/(1-n_{I\sigma}) $.
Then their summation is $g^t_{I\sigma} \tilde{n}_{I\sigma}$. Namely,
all of these terms are proportional to $g^t_{I\sigma}$.
(iii) For the other terms, intersite contractions are taken both for
$I\up$ and $I\dn$. Naive evaluation of these terms does
not yield any explicit factor vanishing at half filling, and we
expect that many terms cancel out each other in some way. Instead, to
derive explicit renormalization, let us consider such contractions for
$\langle k's|(1-\hat{n}_{I\up})(1-\hat{n}_{I\dn})|ks\rangle$, which is
equivalent to $V_{k',k}$ for $k\ne k'$.  Then we can replace as
\begin{equation}
\frac{(1-\hat{n}_{I\up})(1-\hat{n}_{I\dn})}
{\langle P_I^2 \rangle_0}
\Longrightarrow
 \frac{1-n_I}{(1-n_{I\up})(1-n_{I\dn})}
\tilde{n}_{I\up} \tilde{n}_{I\dn}
,
\end{equation}
i.e., all such terms contain $g^t_{I\sigma}/(1-n_{I\bar\sigma})$ explicitly.
These considerations in (i), (ii) and (iii) above demonstrate that
$V_{k',k}$ contains overall factor $g^t_{I\sigma}$.

\section{Magnetic impurity renormalization}
\label{sec:mag}

Let us consider a simple magnetic impurity (\ref{eq:magimpHam}), i.e.,
local magnetic field is applied at site $I$.
Its renormalization can be easily calculated
by subtraction instead of summation of renormalized $\hat{n}_{I\up}$ and $\hat{n}_{I\dn}$
using formulas above.
Corresponding to (\ref{eq:gtnI}) and (\ref{eq:vkk}),
\begin{eqnarray}
\fl
\frac{\langle \gamma_{\ell'} P
S^z_I
 P \gamma_{\ell} \rangle_0 }
{\sqrt{\langle \gamma_{\ell'} P^2 \gamma_{\ell'} \rangle_0
       \langle \gamma_{\ell} P^2 \gamma_{\ell} \rangle_0} }
&\approx
\frac12\left[
\frac{1-n_{I\up}+n_{I\dn}}{1-n_{I\up}}
\langle \gamma_{\ell'}\tilde{n}_{I\up} \gamma_{\ell}^\dagger \rangle_0
-\frac{1-n_{I\dn}+n_{I\up}}{1-n_{I\dn}}
\langle \gamma_{\ell'}\tilde{n}_{I\dn} \gamma_{\ell}^\dagger \rangle_0
\right]
\label{eq:magreno1} \\
&\longrightarrow
\frac{1}{1-n_{I\sigma}}
\left\langle \gamma_{\ell'}
 S^z_I
\gamma_{\ell}^\dagger \right\rangle_0
\qquad
( n_{\up}=n_{\dn} )
,
\label{eq:magreno}
\end{eqnarray}
The renormalization factor for $n_{I\up}=n_{I\dn}$ is
$(1-n_{I\sigma})^{-1}$, which is the square root of the Gutzwiller
renormalization factor for the exchange interaction.  Namely, in
contrast to the non-magnetic impurity, the magnetic impurity is
strengthened by the strong electron correlation. It also makes a good
contrast with the unrenormalized diagonal matrix element $\langle S^z_I
\rangle=\langle S^z_I \rangle_0$ (to derive this, the limit of
$\lambda_{I\up}-\lambda_{I\dn}\rightarrow 0$ should be taken at the end
starting from $\lambda_{I\up}\neq\lambda_{I\dn}$).

In fact, also for magnetic systems ($n_{I\up}\neq n_{I\dn}$), the
factors appearing in (\ref{eq:magreno1}) are equivalent to those in the
renormalization of the exchange interaction derived in
\cite{Fukushima08}, i.e.,
\begin{equation}
\fl
\langle S^z_i S^z_j \rangle
 \approx
\langle S^z_i \rangle_0 \langle S^z_j \rangle_0
 + \frac14 \sum_{\sigma,\sigma'}\eta_{i\sigma,j\sigma'}
\left(\sigma \frac{1 -2\sigma \langle S^z_i \rangle_0 }{1-n_{i\sigma}}\right)
\left(\sigma'\frac{1-2\sigma' \langle S^z_j \rangle_0}{1-n_{j\sigma'}}\right)
.
\end{equation}
Although it is not explicitly noted in \cite{Fukushima08}, in this
renormalization of the spin interaction, the first term is from onsite
contractions and not renormalized (from diagonal matrix elements of the
spin-$z$ operators), whereas the second term including intersite
contractions is enhanced by the renormalization factor (from the second
order of the transition matrix elements of the spin-$z$ operators).
In fact, as shown in the next section,
charge interaction is also renormalized in a similar manner
although the direction of renormalization is opposite.

\section{Charge interaction renormalization}
\label{sec:chargeint}

The conventional GA \cite{FCZhang88} relates
$\langle\hat{O}\rangle$ to
$\langle\hat{O}\rangle_0$ for an operator $\hat{O}$ using a
renormalization factor.
By following this procedure, the renormalization factor is unity for the
charge interaction, namely,
\begin{equation}
\langle \hat{n}_i\hat{n}_j \rangle
\stackrel ? \approx
\langle \hat{n}_i\hat{n}_j \rangle_0
= {n}_i {n}_j + \sum_{\sigma,\sigma'} \eta_{i\sigma,j\sigma'}
\label{eq:convnn}
\end{equation}
However, this approximation is correct only for the leading term ${n}_i
{n}_j$ and the renormalization factor is likely to be derived by taking
only the lowest order into account.
Using a procedure similar to that for the non-magnetic impurity,
more careful analysis can be carried out, i.e.,
\begin{eqnarray}
\fl
&
\langle \hat{n}_i\hat{n}_j \rangle
 \approx
\sum_{\sigma,\sigma'}
\lambda_{i\sigma}\lambda_{j\sigma'}
\frac {\langle
\hat{n}_{i\sigma}(1-\hat{n}_{i\bar\sigma})
\hat{n}_{j\sigma'}(1-\hat{n}_{j\bar\sigma'})
\rangle_0}
{\langle P_i \rangle_0 \langle P_j \rangle_0}
\non \\
\fl & \quad \approx
n_i n_j + \sum_{\sigma,\sigma'}\left(
\eta_{i\sigma,j\sigma'}
- \frac{n_{i\sigma}}{1-n_{i\bar{\sigma}}}\eta_{i\bar{\sigma},j{\sigma'}}
- \frac{n_{j\sigma'}}{1-n_{j\bar{\sigma'}}}\eta_{i\sigma,j\bar{\sigma'}}
+ \frac{n_{i\sigma}}{1-n_{i\bar{\sigma}}}
\frac{n_{j\sigma'}}{1-n_{j\bar{\sigma'}}}\eta_{i\bar{\sigma},j\bar{\sigma'}}
\right)
\non \\
\fl & \quad = n_i n_j +
\sum_{\sigma,\sigma'}g^t_{i\sigma}g^t_{j\sigma'}\eta_{i\sigma,j\sigma'}
.
\label{eq:nnnew}
\end{eqnarray}
At the half filling, any state is an eigenstate of $\hat{n}_i\hat{n}_j$
with the eigenvalue unity by definition because every site is occupied
by one electron and there is no particle number fluctuation, which
contradicts (\ref{eq:convnn}) but is consistent with (\ref{eq:nnnew}).
In fact, the second term of r.h.s.\ of (\ref{eq:nnnew}) is the second
order of (\ref{eq:gtnIgs}), namely, it comes from a process in which
$\hat{n}_j$ creates two quasiparticles and $\hat{n}_i$ annihilates them.

To our knowledge, every calculation in the literature on the GA is using
(\ref{eq:convnn}) instead of (\ref{eq:nnnew}) including the calculation
by the author himself, and probably this error is pointed out for the
first time here.  However, this charge interaction usually does not give
very important contribution in $t$-$J$--type models, and this correction is likely to make only
minor modification to numerical values. Therefore, we expect that main
conclusions are not drastically changed by this correction.
Following this correction, equations in \cite{Fukushima08} should be
modified, namely, $(3 g^s_{ij} - 1)$ and $(3 g^s_{ij} + 1)$ in (14) and
(15) should be replaced by $(3 g^s_{ij} - g^t_{ii}g^t_{jj})$ and $(3
g^s_{ij} + g^t_{ii}g^t_{jj})$, respectively, and derivative of
$g^t_{ii}$ should be also considered for (16).

\section{Conclusion}

Since the Gutzwiller approximation is formulated to (almost) conserve
the particle number at the Gutzwiller projection, one may consider that
quantities related to particle number operators are not
renormalized. However, since the particle number is an expectation value
with respect to an assumed ground state, the constraint of its
conservation does not restrict transition matrix elements with excited
states.
Our results here correct description by the conventional Gutzwiller
approximation in the literature, where such renormalization factors are
calculated with a focus on diagonal matrix elements or lowest-order
terms and regarded as unity.
The results in this paper are general and do not depend on parameters.
Namely, they are valid both for attractive and repulsive impurity
potentials and both for attractive and repulsive charge interactions.

The Fourier-transformed impurity potential has a form of hopping in the
$k$-space. We have derived similarities and differences between this
``hopping'' in the $k$-space and in the real space under real-space
electron double-occupancy prohibition.
As a similarity, they are strongly renormalized to decrease with
hole concentration $x$, and their renormalization factor is
$g^t=2x/(1+x)$ in uniform non-magnetic systems. In addition, the higher
order terms also contain $g^t$.
It should represent that not many available seats to hop are left
because of the electron repulsion.
A difference is, however, $\langle c_{i\sigma}^\dagger c_{j\sigma}
\rangle$ of each $\sigma$ is renormalized in the real space, whereas
renormalization of $\sum_\sigma \langle c_{k'\sigma}^\dagger c_{k\sigma}
\rangle$ appears only after the summation over spin $\sigma=\pm$ in the
$k$-space.
If this summation is replaced by subtraction, which
corresponds to a magnetic impurity in the real space, then the direction
of the renormalization is reversed, i.e., the renormalization factor is
larger than unity and equivalent to the square root of that for the
exchange interaction.
As another difference, the corrections to the leading order term in the
$k$-space can be larger and have more complicated expression than those
in the real space.

As related calculation, renormalization of charge interaction has been
also derived.
The leading order is rather trivial and unrenormalized, i.e., it is the
product of particle densities at the two relevant sites.  The next
leading order term is the second order of transition matrix elements of
the number operators with excited states.  Since the transition matrix
elements are renormalized by $g^t$, these second order terms are
renormalized by $(g^t)^2$, namely, strongly reduced.  These terms
include hopping and pairing amplitude and are relevant to the mean-field
approximation.
Similar relation is found also in the $z$-component of the exchange interaction.
Namely, the leading order is the product of spin-$z$ densities at the
two relevant sites. The next term is the second order of transition
matrix elements of the spin-$z$ operator, which is strengthened
by the electron repulsion.
At the half filling, any state is an eigenstate of $\hat{n}_i\hat{n}_j$,
with the eigenvalue unity. In fact, (\ref{eq:nnnew}) satisfies it even
in magnetic systems, which may demonstrate that the choice of fugacity
factors by (\ref{eq:deflambdai}) is reasonable. Other choices of
fugacity factors also discussed in \cite{Fukushima08} do not seem to
satisfy it in magnetic systems, and their use is likely to be restricted
in systems with small magnetic moments.

\end{document}